\documentclass{aa}

\usepackage{color}
\usepackage{graphicx}
\usepackage{hyperref}
\hypersetup{colorlinks=true,citecolor=blue,linkcolor=blue,anchorcolor=blue}

\usepackage{txfonts}

\begin{document}

   \title{Evolution and star formation history of NGC\,300 from a chemical evolution model with radial gas inflows} 


   \author{Xiaoyu~Kang
          \inst{1,2,3}
          \and
          Rolf-Peter~Kudritzki\inst{4,5}
          \and
          Xiaobo~Gong\inst{6}
          \and
          Fenghui~Zhang\inst{1,2,3}
          }

   \institute{Yunnan Observatories, Chinese Academy of Sciences, 396 Yangfangwang, Guandu District, Kunming, 650216, P.R. China\\
              \email{kxyysl@ynao.ac.cn}
         \and
             Key Laboratory for the Structure and Evolution of Celestial Objects, Chinese Academy of Sciences, 396 Yangfangwang, Guandu District, Kunming, 650216, P. R. China
         \and International Centre of Supernovae, Yunnan Key Laboratory, Kunming 650216, P. R. China
         \and University Observatory Munich,
         Ludwig-Maximilian-Universit$\rm \ddot{a}$t M$\rm \ddot{u}$nchen, Scheinerstr. 1, 81679 Munich, Germany
         \and Institute for Astronomy, University of Hawaii, 2680 Woodlawn Drive, Honolulu, HI96822, USA
         \and School of Opto-electronic Engineering, Zaozhuang University, Zaozhuang, 277160, P.R. China
             }

   \date{Received February 18, 2025; accepted July 14, 2025}

 
  \abstract
   {The cosmic time evolution of the radial structure is one of the key topics in the investigation of disc galaxies. 
   In the build-up of galactic discs gas infall is an important ingredient and it produces radial gas inflows as a physical consequence of angular momentum conservation, since the infalling gas on to the 
   disc at a specific radius has lower angular momentum than 
   the circular motions of the gas at the point of impact.
   NGC\,300 is a well studied isolated, bulge-less, and low-mass disc 
   galaxy ideally suited for an investigation of galaxy evolution
   with radial gas inflows. 
   }
   {Our aim is to investigate the effects of radial gas 
   inflows on the physical properties of NGC\,300, such as, 
   the radial profiles of H{\sc i} gas mass and star formation 
   rate (SFR) surface densities, specific star 
   formation rate (sSFR) and metallicity, as well as to 
   study how the metallicity gradient evolves with cosmic time.}
   {A chemical evolution model for NGC\,300 is constructed by 
   assuming its disc builds up progressively by infalling of 
   metall-free gas and outflowing of metal-enriched gas. 
   Radial gas inflows are also considered in the model. 
   We use the model to build a bridge between 
   the available data (such as gas content, SFR, and chemical 
   abundances) observed today and the galactic key physical 
   processes.}
   {Our model including the radial gas inflows and an inside-out disc 
   formation scenario can simultaneously reproduce the present-day 
   observed radial profiles of H{\sc i} gas mass surface density, SFR surface 
   density, sSFR, gas-phase and stellar metallicity. 
   We find that, although the value of radial gas inflow
   velocity is as low as $-0.1\,{\rm km\,s^{-1}}$, 
   the radial gas inflows 
   steepen the present-day radial profiles of H{\sc i} gas 
   mass surface density, SFR surface density, and 
   metallicity, but flatten the radial sSFR profile. 
   Incorporating radial gas inflows significantly 
   improves the agreement between our model predicted 
   present-day sSFR profile and the observations of 
   NGC\,300.
   Our model predictions are also in good agreement with 
   the star-forming galaxy main sequence and the 
   mass-metallicity relation of star-forming galaxies. 
   It predicts a significant flattening of the metallicity 
   gradient with cosmic time. We also find that the 
   model predicted star formation has been more active recently,
   indicating that the radial gas inflows may be help to 
   sustain star formation in local spirals, at least in NGC\,300.
   }
   {}


   \maketitle
%

\section{Introduction}
\label{sec:intro}

Metallicity acts as a fossil records of evolution and star 
formation history (SFH) of galaxies, because it plays a key role 
in many fundamental galactic physical processes, such as gas 
infall, star formation, stellar evolution, gas outflows.
Metallicity can also provide clues on some additional physical
processes in galaxies, including stellar migration and radial gas 
inflows within galactic discs. The complex interplay between 
the aforementioned physical processes that enhance metal 
production and those that reduce metals in galaxies 
provides important insights on the formation and 
assembly history of galaxies \citep{Sanchez-Almeida2014}.

The study of the metallicity in a galaxy provides an 
essential test bed to explore its disc formation 
scenario and mass assembly history. In particular, 
spiral galaxies in the local Universe universally 
exhibit negative radial metallicity gradients,
with inner regions more metal enriched with respect to the 
outskirts of galactic discs \citep[for example,][]{Zaritsky1994, Magrini2009, Moustakas2010, Stasinska2013, Stanghellini2014, Pilyugin2014, Gazak2015, Zinchenko2019, Liu2022, Bresolin2022, Chenq2023, Kudritzki2024,Sextl2024}.
Although negative radial metallicity gradients are common 
in the local Universe, there is no general consensus yet 
about the behaviour of metallicity gradients cosmic 
evolution: How are the present-day radial metallicity 
gradients established? Do they steepen, flatten, or remain 
fixed with time? In addition, the radial metallicity gradients 
and their temporal evolution encode the scenarios of disc formation, 
reflect the presence of gas infall and outflows, as well as 
radial gas inflows along the disc, and reveal the migration
of stars.

Chemical evolution models, which can build a bridge between 
the chemical abundance patterns observed today and the 
galactic key physical processes, are able to infer the 
SFH of spiral galaxies and the cosmic time 
evolution of their metallicity gradients. 
In this framework, the closed-box chemical evolution model 
\citep{Schmidt1963} failed to explain the relative paucity 
of observed low-metallicity stars (G-dwarf problem) in the 
solar neighbourhood \citep[e.g.,][]{van den Bergh1962, Haywood2019}, 
indicating a necessity for inclusion of continuous gas infall 
in the chemical evolution model \citep{Larson1972, Dalcanton2004}.
To reproduce the observed radial metallicty gradients of 
galactic discs, an "inside-out" disc formation scenario, 
that is, inner regions of disc are formed earlier and 
on a shorter time-scales, has been applied in the model 
and supported by many works \citep[e.g.,][]{Larson1976, Matteucci1989, Chiappini2001, Belfiore2019, Vincenzo2020, Grisoni2018, Frankel2019, Spitoni2021a}.
Low-mass galaxies are more efficient in losing metal-enriched 
matter than high-mass systems because the former have shallower 
gravitational potential wells \citep{Kauffmann1993}. Thus, the 
gas outflow process has become a ubiquitous component 
of galaxy evolution models 
\citep[e.g.,][]{Larson1974, Tremonti2004, Chang2010,Lian2018a, Spitoni2021b, Yin2023}.

In addition, the radial motions of gas and star need to be 
considered in the chemical evolution model.
While extremely difficult to be observed directly,
radial gas flows are postulated
on physical grounds \citep{Lacey1985}. There are several 
mechanisms that could drive such gas flows: i) viscosity 
of the gaseous layer of the disc \citep{Thon1998}, ii) 
gravitational interactions between the gas and the presence 
of the bars or spiral density waves in the disc 
\citep{Kubryk2015}, and iii) mismatch of the angular 
momentum between the infalling gas and the circular 
motions of the gas in the disc 
\citep[e.g.,][]{Portinari2000, Spitoni2011, Grisoni2018, Vincenzo2020, Calura2023}.

Radial migration of stars from their birth place 
to another galaxy region can also affect the metallicity 
gradients. The observed 
age-metallicity relationship in the solar neighbourhood of 
our Galaxy puts strong evidence on the presence of radial 
migration \citep[e.g. ][]{Edvardsson1993, Haywood2008, Schonrich2009,
Feuillet2019, Xiang2022, Lian2022}. In addition, theoretical and numerical studies 
suggest that radial stellar migration can be boosted by 
several process such as mergers or interactions with 
satellites \citep{Quillen2009, Bird2012, Carr2022}, 
the presence of transient spiral structures 
\citep{Sellwood2002, Roskar2008, Daniel2015, Loebman2016}, 
as well as the central bars 
\citep{Minchev2010, Kubryk2013, Halle2015, Khoperskov2020}.

The nearby flocculent low-mass spiral galaxy NGC\,300 is a 
perfect target for studying the secular star formation 
histories and galactic evolution. With a distance of 
$d\,=\,2\,\rm Mpc$ (\citealt{Dalcanton2009},
but see also \citealt{Gieren2005} and \citealt{Sextl2021} for
  slightly different distances), it is the 
closest nearly face-on and isolated \citep{Karachentsev2003}, star-forming \citep{Kruijssen2019} and bulge-less 
\citep{Vlajic2009, Williams2013} disc galaxy with 
well observed radial profiles of gas mass, 
star mass, and star formation rate (SFR) surface densities. 
We will use a chemical evolution model with gas 
infall and outflows to match the observations.


This paper is structured as follows. Section\,\ref{sec:model} 
presents the main ingredients of the chemical evolution model 
used in this work. Section\,\ref{sec:obv} presents the main 
observed data of the target galaxy used to constrain the model.
Section\,\ref{sec:result} presents our results. Sect.\,\ref{sec:sum} 
gives conclusions.


\section{Model}
\label{sec:model}

The chemical evolution model adopted in this work 
is based on \citet{Kang2016}. The
NGC\,300 disc is assumed to progressively  
build up by infalling of primordial gas 
($X\,=\,0.7571, Y_{\rm p}\,=\,0.2429, Z\,=\,0$) 
from its halo and outflowing of metal-enriched gas, 
and it is composed of a number of 
concentric rings. The main improvement of the model is the
implementation of radial inflows of gas following the 
prescriptions described in \citet{Portinari2000} and 
\citet{Spitoni2011}. Radial stellar migration is not 
considered in the model, since the kinematics of globular 
cluster systems \citep{Olsen2004, Nantais2010}, N-body 
simulations studies \citep{Gogarten2010}, and the lack of 
a radial age inversion \citep{Gogarten2010}, as well as 
a pure exponential disc \citep{Bland-Hawthorn2005} 
and the weak transient structure \citep{Cohen2024} 
all together indicate that NGC\,300 has not undergone 
significant radial stellar migrations during its evolution. 

Main ingredients of the model are the inclusion of
infalls of metal-poor gas, star formation law, 
outflows of metal-enriched gas and radial gas 
inflows. The instantaneous recycling approximation (IRA) 
is adopted in the model by assuming 
that stars more massive than $1\rm M_{\odot}$ die 
instantaneously, while those stars less than 
$1\rm M_{\odot}$ live forever. The enriched gas 
is ejected and rapidly becomes well-mixed with 
the surrounding interstellar medium (ISM). IRA 
is an acceptable approximation for chemical 
elements produced by massive stars with short 
lifetimes, such as oxygen. 
On the other hand, IRA is a poor approximation for 
chemical elements produced by stars with long lifetimes, 
such as nitrogen, carbon and iron 
\citep[see][]{Vincenzo2016, Matteucci2021}.
Oxygen is most abundant heavy element by mass in the universe, 
and it is the best proxy for the global metallicity of the galaxy 
ISM. Thus, oxygen abundance (i.e. ${\rm 12+log(O/H)}$) will be 
used to represent the metallicity of NGC\,300 throughout this 
work. The details of the set of equations and 
main ingredients of the model are as follows.

\subsection{Equations of chemical evolution}
\label{sec:equations}

We assume azimuthal homogeneity. The evolution in each
ring at galactocentric radius $r$ during time $t$ can be 
described by three differential equations. The first one 
relates the change of the total mass (stars and gas) 
surface density $\Sigma_{\rm tot}(r,t)$ to the 
rates of gas infall $f_{\rm{in}}(r,t)$ and outflow 
$f_{\rm{out}}(r,t)$ at the corresponding radius and time,
respectively
\begin{equation}
\frac{{\rm d}[\Sigma_{\rm tot}(r,t)]}{{\rm d}t}\,=\,f_{\rm{in}}(r,t)-f_{\rm{out}}(r,t),\\
\label{eq:tot}
\end{equation}
The evolution of the gas mass surface density 
$\Sigma_{\rm gas}(r,t)$ is described in the second 
equation through
\begin{equation}
\frac{{\rm d}[\Sigma_{\rm gas}(r,t)]}{{\rm d}t}\,=\,-(1-R)\Psi(r,t)+f_{\rm{in}}(r,t)-f_{\rm{out}}(r,t)+[\frac{{\rm d}\Sigma_{\rm gas}(r,t)}{{\rm d}t}]_{rf},\\
\label{eq:gas}
\end{equation}
where $\Psi(r,t)$ the SFR surface density at the corresponding 
place and time. $R$ is the fraction of stellar mass returned to 
the ISM. The last term in the equation 
$[\frac{{\rm d}\Sigma_{\rm gas}(r,t)}{{\rm d}t}]_{rf}$ accounts 
for the change of gas mass surface density through the radial 
gas flows and will be described below.

The third differential equation considers the evolution 
of metallicity $Z(r,t)$ as the result of star formation 
and nucleosynthesis. 
It also accounts for the effects of 
radial flows through the term 
$[\frac{{\rm d}[Z(r,t)\Sigma_{\rm gas}(r,t)]}{{\rm d}t}]_{rf}$, 
which will also be described below
\begin{eqnarray}
\frac{{\rm d}[Z(r,t)\Sigma_{\rm gas}(r,t)]}{{\rm d}t}\,=\,y(1-R)\Psi(r,t)-Z(r,t)(1-R)\Psi(r,t) \nonumber\\
+Z_{\rm{in}}f_{\rm{in}}(r,t)-Z_{\rm{out}}(r,t)f_{\rm{out}}(r,t) \nonumber\\
+[\frac{{\rm d}[Z(r,t)\Sigma_{\rm gas}(r,t)]}{{\rm d}t}]_{rf},
\label{eq:metallicity}
\end{eqnarray}
where $y$ is the nuclear metal synthesis yield.

For the values of $R$ and $y$, we adopt 
$R\,=\,0.289$ and $y\,=\,0.0099$, which are averages 
over metallicity obtained from Table\,2 of 
\citet{Vincenzo2016} (see also \citealt{Romano2010}) for
the intial mass function (IMF)  of \citet{Kroupa1993}. 
The dependence on metallicity and time is weak and, thus, 
we work with constant average values. On the other hand, 
the chemical enrichment of galaxies depends largely on 
the IMF \citep{Vincenzo2016, Goswami2021} and the Kroupa 
IMF is the preferred describing the chemical evolution 
of spiral discs as pointed out in the references just 
given. 

$Z_{\rm{in}}$ is the metallicity of the infalling 
gas. In this work, $Z_{\rm{in}}$ is assumed to be a function 
of time to approximate the recycling effects caused by mixing 
of the infalling gas from intergalactic medium (IGM) with 
gas outflows in the circumgalactic medium (CGM) of NGC\,300, 
after which gas is recycled back into the disc.
\citet{Kang2023} found that most of the stellar mass 
of M33-mass bulgeless spiral galaxies were assembled at 
$z\,<\,1$. As a result, 
$Z_{\rm{in}}$ is assumed to be primordial ($Z_{\rm{in}}=0$) 
at redshift $z\,>\,0.7$. For $z\,\leq\,0.7$, we model 
its metallicity as a time dependent function to 
approximate the recycling effect, i.e., 
$Z_{\rm{in}}/{\rm Z_{\odot}}=k\times\,t+b$, where 
$t$ represents the evolutionary time in units of 
$\rm Gyr$, and the age of the universe at $z\,=\,0$ 
is $13.5\,\rm Gyr$.
The slope of the time-dependent metallicity 
function ($k$) for the infalling gas is determined 
by referencing the metallicity evolution trend in the 
NGC\,300 disc, as shown in the lower-right panel of 
Figure\,8 of \citet{Kang2016}. 
The $Z_{\rm{in}}/{\rm Z_{\odot}}$-intercept ($b$) 
is obtained from the metallicity of CGM at the present 
day, which spans $-1.0<{\rm log}(Z/{\rm Z_{\odot}})<-0.5$ 
\citep{Prochaska2017, Pointon2019}. As for 
NGC\,300, the present-day metallicity of CGM is 
assumed to be $-0.75$. Therefore, the functional 
form of metallicity of the infalling gas is given by 
$Z_{\rm{in}}/{\rm Z_{\odot}}=0.039\times\,t-0.348$.

$Z_{\rm{out}}(r,t)$ is the metallicity of the outflowing gas, 
and its value is assumed to be identical to 
that of ISM at the time the outflows are launched, 
i.e. $Z_{\rm{out}}(r,t)=Z(r,t)$ \citep{Ho2015,Kudritzki2015}.


\subsection{Main ingredients}
\label{sec:ingrdients}

Continuous infall of primordial gas is invoked in 
the chemical evolution model to explain the relative scarcity 
of observed low-metallicity stars (G-dwarf problem) in galaxy discs 
\citep[e.g.,][]{van den Bergh1962, Haywood2019}.
The primordial gas infall rate at each radius $r$ and time $t$,
$f_{\rm in}(r, t)$, in units of $\rm M_\odot\,pc^{-2}\,Gyr^{-1}$,
is expressed by:
\begin{equation}
f_{\rm{in}}(r,t)=A(r)\cdot t\cdot e^{-t/\tau},
\label{eq:infall rate}
\end{equation}
where $\tau$ is the gas infall timescale. Following \citet{Matteucci1989}
and \citet{Kang2016}, the infall timescale
has the form $\tau(r)\,=\,a\times r/{R_{\rm d}}+b$.
In this way, it takes the gas longer to settle onto
the disc in the outer regions, corresponding to a scenario 
inside-out disc formation. $R_{\rm d}\,=\,1.29\,\rm kpc$ is 
the present-day disc scale-length and derived from the 
observed $K-$band luminosity distribution \citep{MM2007} 
together with a total stellar mass of 
$M_{*}\,=\,1.928\times10^{9}~\rm M_{\odot}$ \citep{MM2007}.
The coefficients $a$ and $b$ for $\tau(r)$ are free parameters 
in our model and will be determined below.

While a simple exponential form of gas infall rate 
is a popular assumption in many previous chemical evolution
models \citep[e.g.,][]{Matteucci1989, Yin2009, Spitoni2020, Calura2023}, the gas infall rate form adopted in this work 
includes more possible scenarios and is more suitable for 
low-mass galaxies. The reader is referred to \citet{Kang2016} 
for a more indepth descriptions of the difference between these 
two gas infall forms.

\begin{figure}
  \centering
  \includegraphics[angle=0,width=0.475\textwidth]{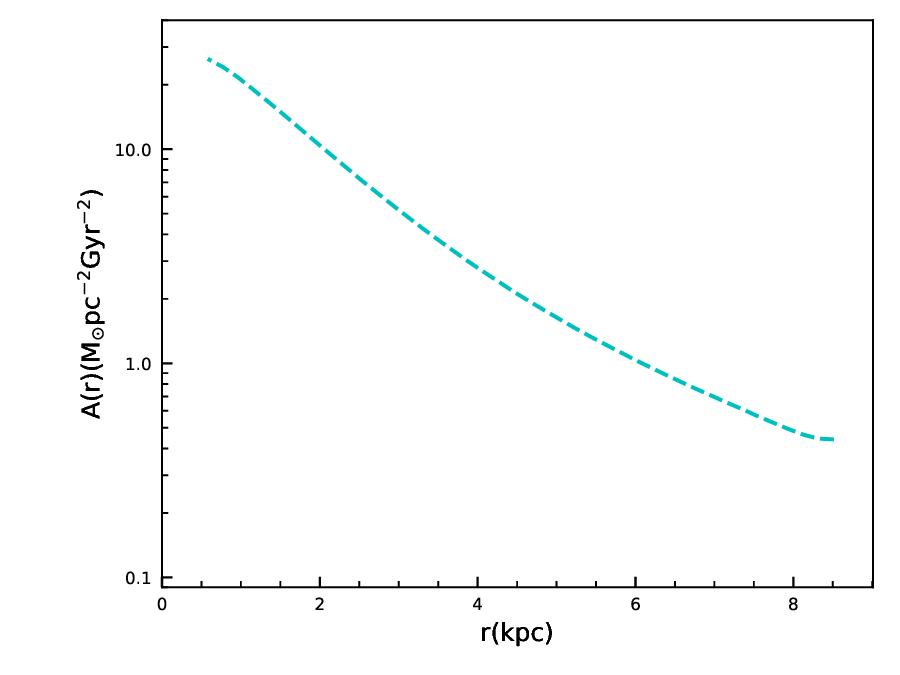}
    \caption{The function $A(r)$ obtained with the best-fitting 
    model of NGC\,300 (see Section\,\ref{sec:result}). 
    }
  \label{Fig:A_r}
\end{figure}

The function $A(r)$ modifies the infall rate as a function of
galactocentric radius and is iteratively constrained by requesting
that the present-day model stellar mass surface density  
$\Sigma_*(r,t_{\rm g})$ follows the observed  exponential profile:
\begin{equation}
\Sigma_*(r,t_{\rm g})=\Sigma_*(0,t_{\rm g}){\rm exp}(-r/R_{\rm d}),
\label{eq:stellar_distrbution}
\end{equation}
$\Sigma_*(0,t_{\rm g})$ is the present-day central stellar mass
surface density, and it can be obtained from
$\Sigma_*(0,t_{\rm g})=M_{*}/2\pi R_{\rm d}^{2}$. 
$t_{\rm g}$ is the cosmic age and with $t_{\rm g}=13.5\rm\,Gyr$ 
according to the standard flat cosmology.

In our calculation, after fixing the values of $\tau$, 
we start with an initial distributioon of $A(r)$ and numerically
solve the gas evolution (Eq.\,\ref{eq:gas}) and 
the increase of stellar mass (via Eq.\,\ref{eq:tot}) 
adopting a SFR surface density $\Psi(r,t)$ (see below). 
By comparing the resulted $\Sigma_*(r,t_{\rm g})$ with its
observed value, we adjust the value of $A(r)$ and repeat
the calculation until the resulting $\Sigma_*(r,t_{\rm g})$
agrees well with the observed radial distribution. Figure\,\ref{Fig:A_r} plots the best-fitting model 
(see Section\,\ref{sec:result}) predicted 
radial profile of $A(r)$.

The SFR surface density $\Psi(r,t)$ (in units of
$\rm{M_{\odot}}\,{pc}^{-2}\,{Gyr}^{-1}$) describes 
the amount cold gas turning into stars per unit time. 
Following \citet{Leroy2008}, \citet{Krumholz2014} and
\citet{Kang2016,Kang2023}, we adopt $\Psi(r,t)$ proportional 
to the molecular gas mass surface density 
$\Sigma_{\rm H_2}(r,t)$ 
\begin{equation}
\Psi(r,t)=\Sigma_{\rm{H_2}}(r,t)/t_{\rm dep},
\label{eq:h2sfr}
\end{equation}
$t_{\rm dep}$ is the molecular gas depletion time, 
and its value is taken as $t_{\rm dep}\,=\,1.9\,\rm Gyr$ 
\citep{Leroy2008}. In order to calculate $\Sigma_{\rm H_2}(r,t)$, 
we need to split the total gas mass surface density into 
its atomic and molecular components. This is done in the 
same way as described in detail in \citet{Kang2023}. 

NGC\,300 is a low-mass disc galaxy with shallow 
gravitational potential and a low rotation speed,
making it efficient in expelling metal-enriched matter
\citep{Tremonti2004, Hirschmann2016, Lian2018a, Spitoni2020}. 
The gas outflow rate $f_{\rm out}(r,t)$ (in units of 
$\rm{M_{\odot}}\,{pc}^{-2}\,{Gyr}^{-1}$) is assumed to be 
proportional to $\Psi(r,t)$ \citep[see ][]{Recchi2008}:
\begin{equation}
f_{\rm out}(r,t)=b_{\rm out}\Psi(r,t),
\label{eq:outflow}
\end{equation}
where $b_{\rm out}$ is the gas outflow efficiency 
(dimensionless quantity), and it is also a free
parameter in the model. 
We emphasize that the outflow efficiency is 
assumed to be constant, since the outflow process is 
energetically driven by star formation, and 
$b_{\rm out}$ simply reflects the efficiency of 
energy transfer from star formation to the gas outflow.

\subsection{Implementation of radial inflows}
\label{sec:radial_gas inflow}

The groundbreaking works of \citet{Tinsley1978} and 
\citet{Mayor1981} highlight the potential importance 
of radial gas flows for the chemical evolution of 
galactic discs. Radial gas flows are the physical 
consequence of the gas infall, because the specific 
angular momentum of the infalling gas is lower than the 
gas circular motions in the disc, and mixing between the 
two will induce a net radial gas inflow. In consequence,
they should be considered if one assumes that 
the galactic disc is formed by gas infall. 
We implement the radial gas inflows in our chemical evolution 
model based on the prescriptions and formalism of 
\citet{Portinari2000}and \citet{Spitoni2011}. 
Following \citet{Spitoni2013} and \citet{Grisoni2018}
the radial inflow velocity is assumed to be 
related to the galactocentric distance, i.e., 
$\mid\upsilon_{R}\mid\,=\,c\times r/{R_{\rm d}}+d$, where $c$ and 
$d$ are the coefficients for $\upsilon_{R}$.

In our numerical solution of equations (1) to (3) we divide 
the galactic disc into discrete shells. The $k$-th shell has the
galactocentric radius $r_{k}$ and inner and outer edges named as 
$r_{k-\frac{1}{2}}$ and $r_{k+\frac{1}{2}}$, respectively. 
Through these edges, gas inflow can occur with velocities 
$\upsilon_{k-\frac{1}{2}}$ and $\upsilon_{k+\frac{1}{2}}$, 
respectively. The gas flow velocities are taken positive 
outwards and negative inwards.

Radial inflows through the edges, with a flux $F(r)$, alter 
the gas mass surface density $\Sigma_{\rm gas}(r_{k})$ in the 
$k$-th shell according to: 
\begin{equation}
[\frac{{\rm d}\Sigma_{\rm gas}(r_{k})}{{\rm d}t}]_{rf}\,=\,-\frac{1}{\pi(r_{k+\frac{1}{2}}^{2}-r_{k-\frac{1}{2}}^{2})}[F(r_{k+\frac{1}{2}})-F(r_{k-\frac{1}{2}})],\\
\label{eq:radial_gas}
\end{equation}
where the gas flow at $r_{k-\frac{1}{2}}$ and $r_{k+\frac{1}{2}}$ 
can be written as
\begin{equation}
F(r_{k-\frac{1}{2}})\,=\,2\pi r_{k-\frac{1}{2}}\upsilon_{k-\frac{1}{2}}[\Sigma_{\rm gas}(r_{k-1})],\\
\label{eq:flow_minus}
\end{equation}
and 
\begin{equation}
F(r_{k+\frac{1}{2}})\,=\,2\pi r_{k+\frac{1}{2}}\upsilon_{k+\frac{1}{2}}[\Sigma_{\rm gas}(r_{k+1})],\\
\label{eq:flow_add}
\end{equation}
The inner edge of $k$-th shell, $r_{k-\frac{1}{2}}$, is taken at 
the mid-point between the characteristic radii of the shells $k$-th 
and $k-1$-th, 
\begin{equation}
r_{k-\frac{1}{2}}\,=\,(r_{k-1}+r_{k})/2,\\
\label{eq:radius_minus}
\end{equation}
and similarly for the outer edge $r_{k+\frac{1}{2}}$
\begin{equation}
r_{k+\frac{1}{2}}\,=\,(r_{k}+r_{k+1})/2,\\
\label{eq:radius_add}
\end{equation}
From Eqs.\,\ref{eq:radius_minus} and \ref{eq:radius_add}, we can obtain that 
\begin{equation}
(r_{k+\frac{1}{2}}^{2}-r_{k-\frac{1}{2}}^{2})\,=\,\frac{r_{k+1}-r_{k-1}}{2}(r_{k}+\frac{r_{k-1}+r_{k+1}}{2}),\\
\label{eq:square}
\end{equation}
Combining Eqs.\,\ref{eq:radial_gas}, \ref{eq:flow_minus}, 
\ref{eq:flow_add}, and \ref{eq:square} 
and , we can get radial flow term 
$[\frac{{\rm d}[Z(r_k,t)\Sigma_{\rm gas}(r_{k},t)]}{{\rm d}t}]_{rf}$ of Eq.\ref{eq:metallicity} as follows:
\begin{eqnarray}
[\frac{{\rm d}[Z(r_k,t)\Sigma_{\rm gas}(r_{k},t)]}{{\rm d}t}]_{rf}\,=\,
-\beta_{k}Z(r_{k},t)\Sigma_{\rm gas}(r_{k},t) \nonumber\\
+\gamma_{k}Z(r_{k+1},t)\Sigma_{\rm gas}(r_{k+1},t),
\label{eq:flow_term}
\end{eqnarray}
where 
\begin{equation}
\beta_{k}\,=\,-\frac{2}{r_{k}+\frac{r_{k-1}+r_{k+1}}{2}}\times[\upsilon_{k-\frac{1}{2}}\frac{r_{k-1}+r_{k}}{r_{k+1}-r_{k-1}}],\\
\label{eq:beta}
\end{equation}
and
\begin{equation}
\gamma_{k}\,=\,-\frac{2}{r_{k}+\frac{r_{k-1}+r_{k+1}}{2}}\times[\upsilon_{k+\frac{1}{2}}\frac{r_{k}+r_{k+1}}{r_{k+1}-r_{k-1}}].\\
\label{eq:gama}
\end{equation}

In summary, we note that our model has five free parameters 
($a$, $b$, $c$, $d$, and $b_{\rm out}$). The first two - 
$a$ and $b$ -  characterize the gas infall time scale 
$\tau$, the next two - $c$ and $d$ - describe the radial gas 
flow within the disc, and the fifth parameter $b_{\rm out}$ 
constrains the gas outflow efficiency. The determination 
of the best combination of these parameters will be 
described in Sect.\,\ref{sec:result}.

\section{Observations}
\label{sec:obv}
The major goal of our chemical evolution model is to 
use the stellar and gas content, SFR, and chemical abundances 
observed today to infer the galaxy star formation history 
(SFH) and how the chemical composition and metallicity gradients evolve with cosmic time. 
The details of the available observations used to constrain 
the model are described in the following.

The atomic hydrogen (H{\sc i}) gas of NGC\,300 is obtained 
from the Australia Telescope Compact Array (ATCA), and 
the radial profile of the H{\sc i} mass surface density
($\Sigma_{\rm HI}$) of NGC\,300 is taken from \citet{Westmeier2011}. 
As already mentioned above, the stellar mass of the NGC\,300
disc is obtained from the $K-$band luminosity. The value is
$M_{\rm \ast}\sim 1.928\times10^{9}\rm M_{\odot}$ 
\citep{MM2007}.

The radial profiles of SFR surface density ($\Sigma_{\rm SFR}$) 
of NGC\,300 is determined from Hubble Space Telescope(HST) 
resolved stellar populations \citep{Gogarten2010}, the 
combination of far-ultraviolet (FUV) and $24\,\mu\rm m$ maps 
\citep{Williams2013}, and the FUV image \citep{Mondal2019}.
The current total SFR in the disc of NGC\,300 is reported 
to be in the range $\sim0.08-0.46\rm M_{\odot }$\,yr$^{-1}$ 
using different tracers, such as the X-ray luminosity 
\citep{Binder2012}, FUV luminosity \citep{Karachentsev2013, Mondal2019}, 
H${\alpha}$ emission \citep{Helou2004, Karachentsev2013, Kruijssen2019}, 
mid-infrared \citep[MIR,][]{Helou2004}, the HST 
resolved stars \citep{Gogarten2010} and the 
spectral energy distribution (SED) modeling 
\citep{Casasola2022, Binder2024}.

The specific SFR (sSFR) is defined as SFR per unit of stellar mass.
The observed sSFR value for NGC\,300 can be computed from  
${\rm sSFR}\,=\,\frac{\rm SFR}{M_{*}}$, 
i.e., $\rm -10.382\,\leq\,log(sSFR/yr^{-1})\,\leq\,-9.622$.
The radial profile of sSFR has been derived by \citet{MM2007}, 
who used the (FUV$-K$) color profile of NGC\,300 and 
adopted a proper SFR calibration of the FUV 
luminosity and $K-$band mass-to-light ration.

Crucial additional constraints on the evolution of NGC 300 are
obtained from the radial distribution of metallicity. Contrary to
our previous work \citep{Kang2016, Kang2023}, we additionally 
rely on accurate metallicity
measurements based on  detailed non-LTE spectroscopy of very 
young massive stars, blue supergiants \citep[BSGs,][]{Kudritzki2008} and red supergiants \citep[RSGs,][]{Gazak2015}. (Note that we 
have applied the relationship in the appendix of 
\citealt{Davies2017} to put the RSG on the same metallicity scale 
as the BSG). We convert the metallicities of these very young 
stars to oxygen abundances via 12 + log(O/H) = log
Z/Z$_{\odot}$ +8.69 using the solar oxygen abundance
\citep{Asplund2009}. We then also use the H{\sc ii} regions 
oxygen abundances obtained by \citet{Bresolin2009}, which are
based on the analysis of collisionally excited lines and 
the T$_{\rm e}$ method. In addition, we also compare with 
oxygen abundances of planetary nebulae (PNe) \citep{Stasinska2013}, 
which represent metallicities at intermediate ages.

\section{Results and discussion}
\label{sec:result}

\begin{figure*}
  \centering
  \includegraphics[angle=0,scale=0.9]{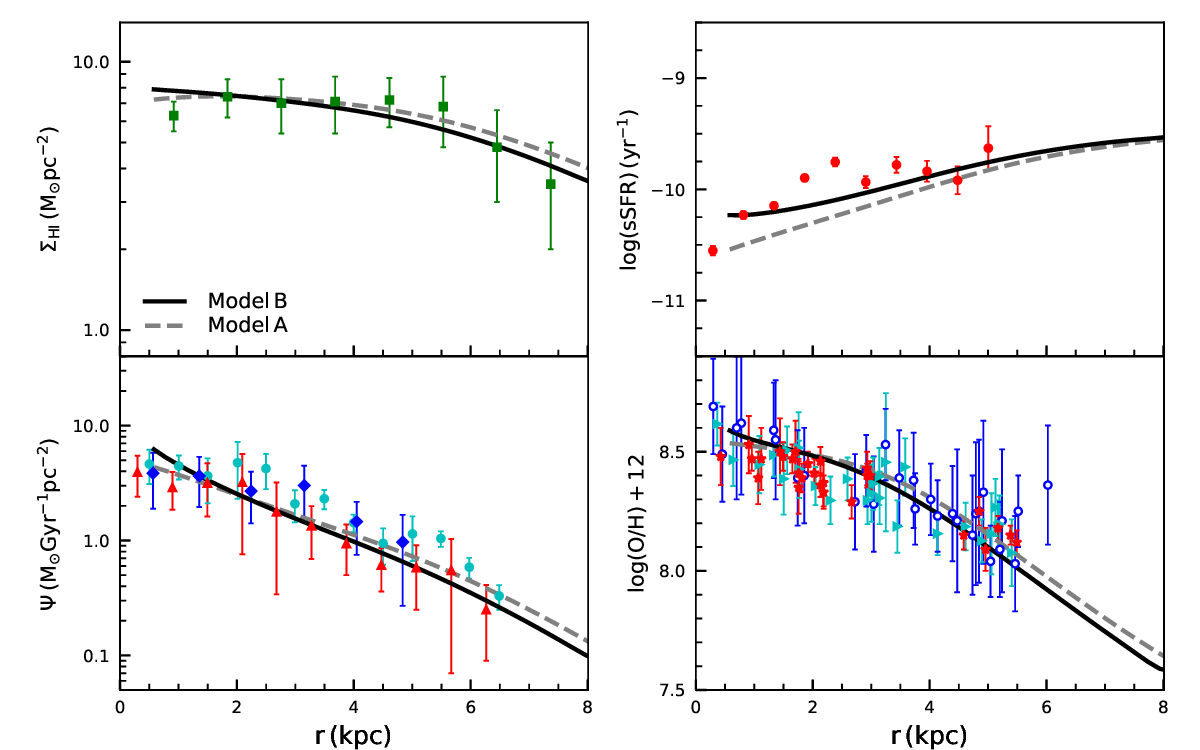}
    \caption{Comparisons between the model predictions and the 
    observed data of NGC\,300. Solid lines correspond to the best-fitting 
    model with a radial inflow of gas (Model\,B) and dashed lines to the  model without radial inflow 
    (Model A). The left-hand side shows the radial profiles of H{\sc i} 
   (top) mass and SFR (bottom) surface density, while the 
   right-hand side displays the radial profiles of sSFR (top) 
   and $\rm 12+log(O/H)$ (bottom). H{\sc i} data (see text) are 
   shown by green filled squares. SFR data are separately 
   denoted as blue filled diamonds \citep{Gogarten2010}, red 
   filled triangles \citep{Williams2013}, and cyan filled 
   cycles \citet{Mondal2019}. sSFR data (see text) are plotted 
   by red filled cycles. ${\rm 12+log(O/H)}$ data from H{\sc ii}
   regions \citep{Bresolin2009}, BSGs \citep{Kudritzki2008}, and 
   RSGs \citep{Gazak2015} are shown as red filled asterisks, blue 
   open circles, and cyan filled triangles, respectively.
   }
\label{fig:model_result}
\end{figure*}

\begin{figure}
\includegraphics[angle=0,scale=0.55]{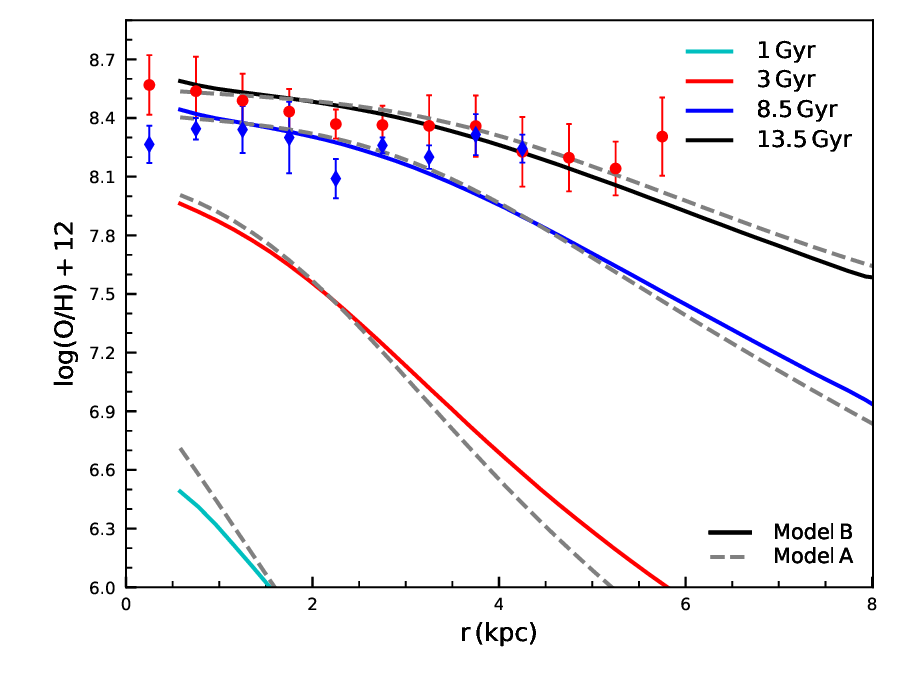}
   \caption{Cosmic time evolution of metallicity radial profiles.
   Solid line with different colours represent Model\,B at 
   $1\,\rm Gyr$ (cyan), $3\,\rm Gyr$ (red), $8.5\,\rm Gyr$ 
   (blue), and $13.5\,\rm Gyr$ (present-day, black). 
   The corresponding results predicted by Model\,A 
   are shown as gray dashed lines. Red solid cycles are 
   the binned metallicity for H{\sc ii} regions from \citet{Bresolin2009}, BSGs from \citet{Kudritzki2008}, 
   and RSGs from \citet{Gazak2015}, while blue solid diamonds 
   denote the binned metallicity of PNe from \citet{Stasinska2013}.
   }
\label{fig:dif_r_Zgas}
\end{figure}

The collection of observed data displayed in
Figure.\,\ref{fig:model_result} is used to constrain our
model and its five free parameters (see Sect.\ref{sec:model}). 
We firstly computed the radial profiles of H{\sc i} 
mass and SFR surface density, sSFR, and $\rm 12+log(O/H)$ along 
the disc of NGC\,300 and compare them to available observed data 
to search for the best-fitting model for NGC\,300. Note that the $\rm
12+log(O/H)$ data are all together combined into the 12 radial bins
displayed in  Fig.\,\ref{fig:dif_r_Zgas}.

As in \citet{Kang2016}, the classical $\chi^{2}$ technique 
is adopted to calculate the best combination of free parameters, 
which we vary within the range of 
$0<\,a\,\leq3.0$, $1.0\leq\,b\leq\,5.0$, 
$0.1\leq\,\,0\leq c\times r/{R_{\rm d}}+d\leq1.0$, and 
$0<\,b_{\rm out}\leq\,1.0$. 
We obtain ($a$, $b$, $c$, $d$, $b_{\rm out}$)\,=\,
(0.16, 3.0, 0.0, 0.1, 0.6), i.e., 
$(\tau,\upsilon_{R}, b_{\rm out})\,=\,(0.16r/{R_{\rm d}}+3.0\,{\rm Gyr},-0.1\,{\rm km\,s^{-1}}, 0.6)$. 
The model calculated with these parameters is 
called Model\,B. The results corresponding to 
Model\,B are plotted by solid lines in 
Fig.\,\ref{fig:model_result}. 
A remarkable agreement is found between the Model\,B predictions 
and the NGC\,300 observations, that is, the Model\,B results 
can simultaneously reproduce the radial observed profiles of 
H{\sc i} gas mass surface density, SFR surface density, 
sSFR and $\rm 12+log(O/H)$. 

The best-fitting model of NGC\,300 without radial gas 
inflows in \citet{Kang2016} (``Model\,A'') with parameters
$(\tau, \upsilon_{R}, b_{\rm out})=(0.52r/{\rm r_{d}}+2.6\,{\rm Gyr}, 0, 0.9)$ is also plotted as in Fig.\,\ref{fig:model_result} as 
a dashed line. We see that Model\,A is also basically able 
to reproduce the observed data. The differences between the 
two models are small, but there are clear systematic trends. 
The radial gas inflows steepen the present-day radial profiles 
of H{\sc i} gas mass surface density, SFR surface density, 
and metallicity, but flatten the radial sSFR profile. 
The steeper profiles of gas mass surface density and 
metallicity predicted by Model\,B are consistent with the 
result of \citet{Calura2023}. We also note that the central 
increase in metallicity of Model B is in better agreement 
with the observations.

$\upsilon_{R}\,=\,-0.1\,{\rm km\,s^{-1}}$ appears to be a very small 
value. However, it is supported by a study of 54 local spiral 
galaxies based on high-sensitive and high-resolution data 
of the H{\sc i} emission line, which finds that the
radial inflow velocities are generally small, 
with an average inflow rate of about $-0.3\,{\rm km\,s^{-1}}$  \citep{Di Teodoro2021}.

\begin{figure*}
  \centering
  \includegraphics[angle=0,scale=0.55]{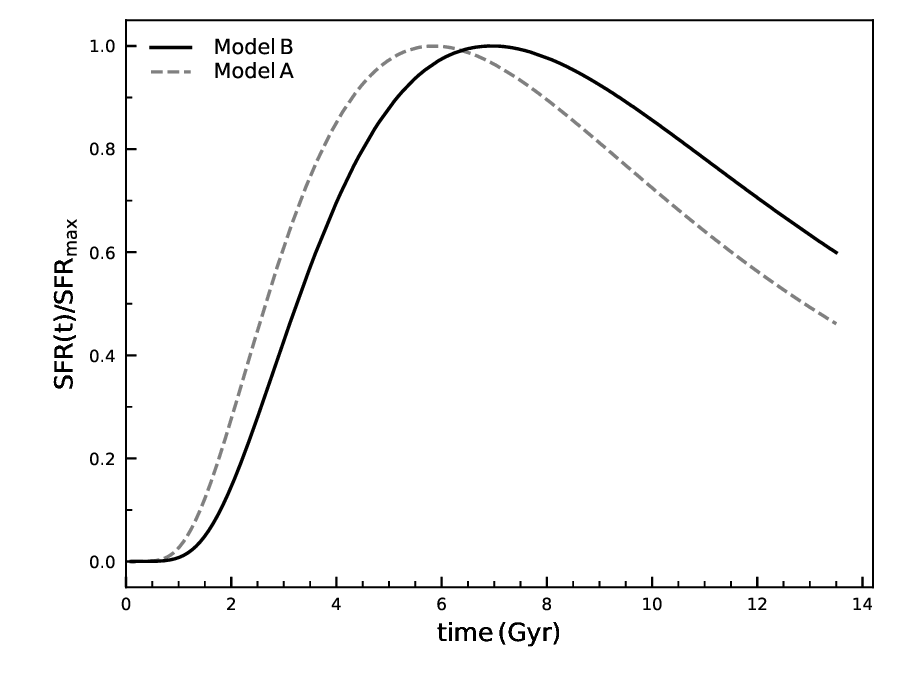}
  \includegraphics[angle=0,scale=0.55]{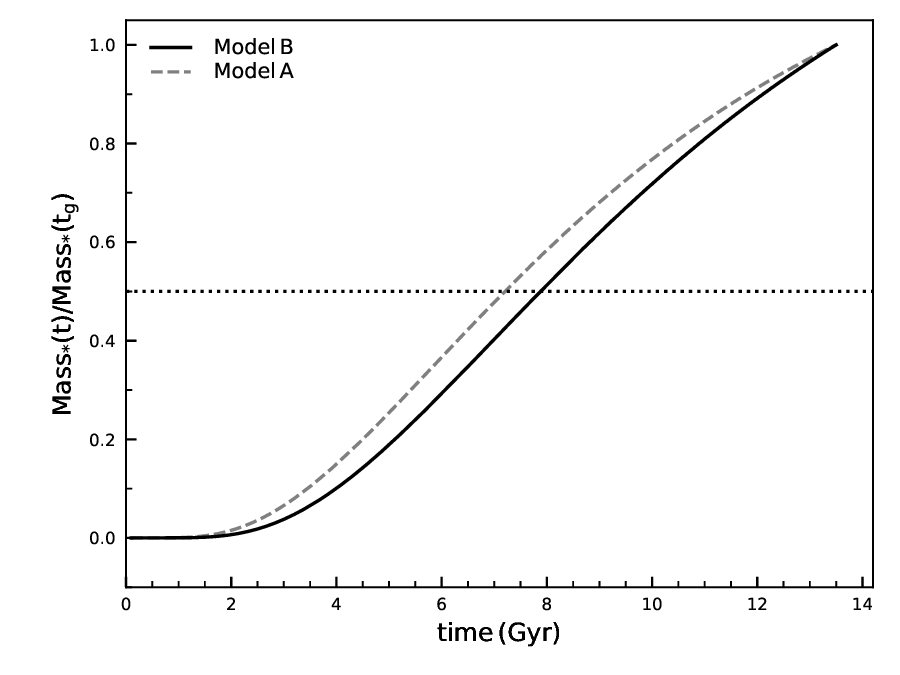}
    \caption{SFH (left) and stellar mass 
    growth history (right) of NGC\,300 predicted by Model\,A 
    (dashed line) and Model\,B (solid line). SFH is normalized 
    to its maximum value, while stellar mass is normalized to 
    its present-day value. The dotted line in the right panel 
    denotes when the stellar mass achieves 50\% of its final 
    value.
    }
  \label{fig:SFR_star_t}
\end{figure*}

The left panel of Fig.\,\ref{fig:SFR_star_t} shows that, 
compared to Model\,A, Model\,B predicts both delayed 
and more extended star formation.
The peak of SFR from Model\,B is shifted by 1\,Gyr. 
The SFR predicted by Model\,B and Model\,A are respectively 
$\sim0.16\rm M_{\odot }$\,yr$^{-1}$ and 
$\sim0.15\rm M_{\odot }$\,yr$^{-1}$, consistent with 
the observed values (see Section\,\ref{sec:obv}).
The right panel of Fig.\,\ref{fig:SFR_star_t} shows that, 
both Model\,A and Model\,B predicted galaxy stellar masses 
have been steadily increasing to their present-day values. 
Compared to Model\,A, Model\,B forms stars later.
The right panel agrees well with the bottom panel of Fig\,17 
in \citet{Sextl2023}, which based on the galaxy evolution model 
of \citet{Kudritzki2021a} led to the conclusion that most 
of galaxy stellar masses has been assembled in the last 
$10\,\rm Gyr$. Fig.\,\ref{fig:SFR_star_t} indicates that 
the radial gas inflows help to sustain star formation 
in local spirals, at least in NGC\,300.

\begin{figure}
  \centering
  \includegraphics[angle=0,scale=0.55]{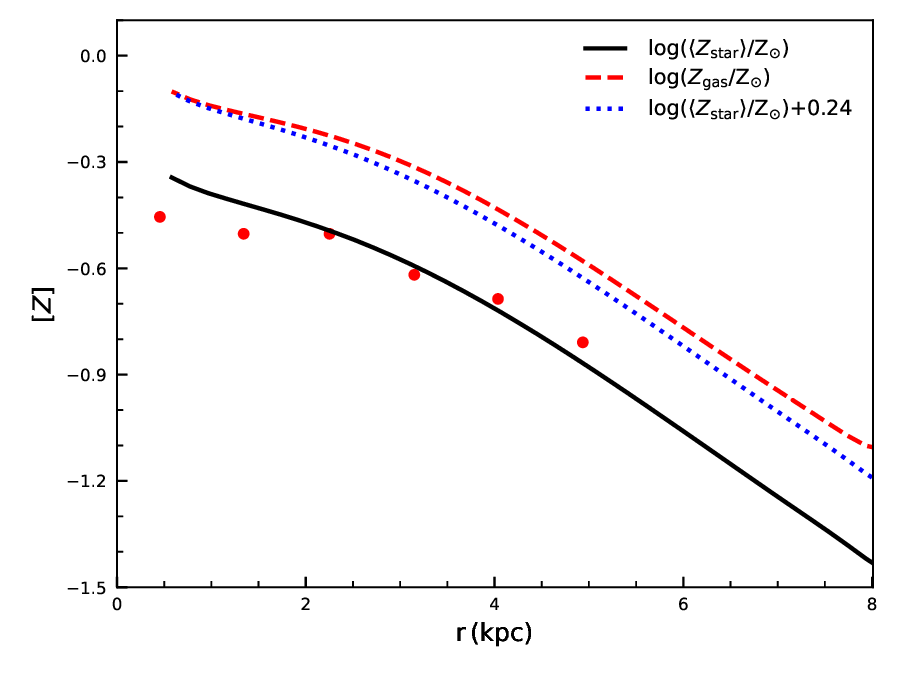}
  \caption{Radial distribution of the mean metallicity of the
    entire stellar population. Observations obtained from the
    photometric analysis of color-maginitude diagrams
    \citep[CMD,][]{Gogarten2010} are shown as red circles 
    and the Model\,B prediction is displayed as the black solid 
    line. The Model\,B predicted metallicity of the ISM and the 
    very young stars is additionally shown as the red dashed 
    curve. For comparison purposes, the blue dotted curve 
    represents the black curve shifted by +0.24\,dex.
}
\label{fig:stellar-gas-Z}
\end{figure}

An additional test of our model is obtained by a comparison with 
the observed mean metallicities for the entire stellar population 
as derived from the photometric analysis of color-magnitude 
diagrams (CMD) in \citet{Gogarten2010}. The comparison is 
carried out in Fig.\,\ref{fig:stellar-gas-Z}. The model 
mass-weighted average stellar metallicity at galactocentric
radius $r$ and time $t$ is calculated as \citep{Pagel1997, Kang2021}
\begin{equation}
\langle\,Z_{\rm star}(r, t_{\rm g})\rangle\,=\,\frac{\int^{t_{\rm g}}_{0} Z_{\rm gas}(r,t'){\rm SFR}(r,t'){\rm d}t'}{\int^{t_{\rm g}}_{0} {\rm SFR}(r,t'){\rm d}t'}.
\label{eq:Zstar}
\end{equation}
Given the uncertainties of the purely photometric 
diagnostics the agreement is quite compelling. 
We note the Model\,B predicted metallicity 
average over all different stellar ages is about 
$\sim0.2\,\rm dex$ smaller than the one of
the present-day ISM and the very young stellar population. 
This agrees with cosmological simulations and corresponding 
chemical evolution models 
\citep[][etc.]{Finlator2008, Pipino2014, Peng2014} and 
the observational work
\citep[e.g.,][]{Halliday2008,Fraser-McKelvie2022}. 
Based on analyzing the integrated stellar population 
absorption line spectra of $\sim\,200,000$ star-forming 
galaxies in the Sloan Digital Sky Survey, \citet{Sextl2023} 
also derived that, for a galaxy with stellar mass 
(${\rm log}(M_{\ast}/{\rm M}_{\odot})\,\sim\,9.3$), 
the metallicity difference between the young stellar 
population and average metallicity is about $\sim0.2\,\rm dex$.
In addition, we find that the Model\,B predicted 
stellar metallicity 
gradient exhibits a steeper slope compared to the gas-phase 
gradient, which aligns with previous statistical findings 
from observational studies
\citep[][and reference therein]{Lian2018b, Sanchez-Blazquez2020}.

\begin{figure}
  \centering
  \includegraphics[angle=0,scale=0.55]{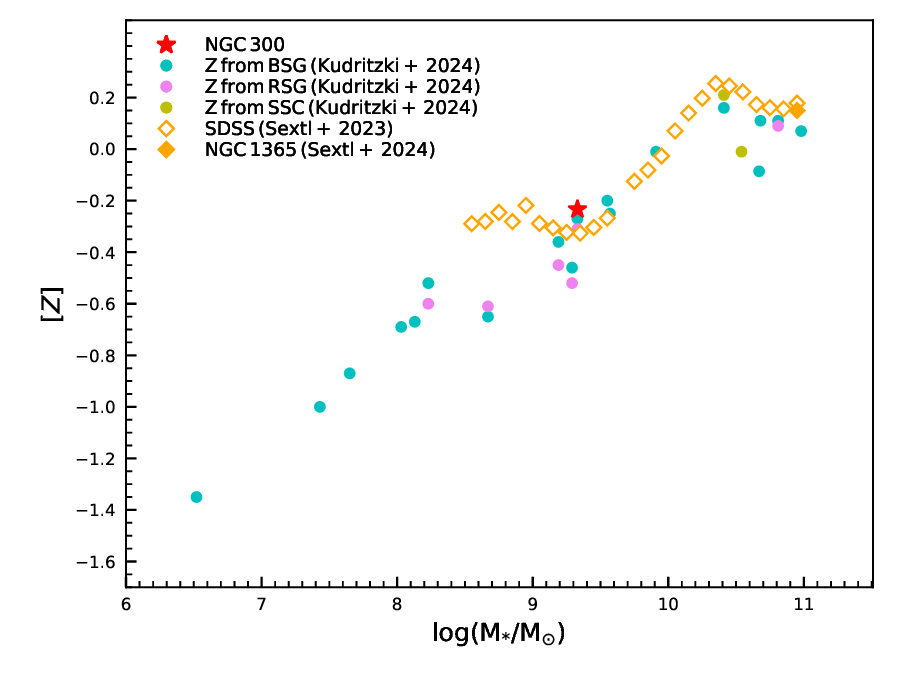}
    \caption{The mass-metallicity relationship of star-forming galaxies. The prediction by Model\,B shown as a red solid 
    star is compared with observations of a large sample of 
    galaxies. Solid circles refer to stellar metallicity 
    derived from spectroscopy of individual blue
    supergiants (cyan), red supergiants (pink), and super star 
    clusters (yellow) \citep[see][and references therein]{Kudritzki2024}. The open orange diamonds represent 
    the result of integrated galaxy spectra of young stellar population for 250,000 SDSS star-forming 
    galaxies by using a stellar population synthesis technique
    \citep{Sextl2023}, while the orange solid diamonds denotes the 
    analysis of the spatially resolved young stellar population 
    synthesis study \citep{Sextl2024}.
}
\label{fig:M_Z}
\end{figure}

An additional important test of our galaxy evolution model is
the prediction with respect to the mass-metallicity relationship
(MZR) of star-forming galaxies. This is a key result of 
a comprehensive spectroscopic project which revealed a 
tight relationship between the metallicity of the young 
stellar population and total galaxy stellar mass. 
Fig.\,\ref{fig:M_Z} displays the latest results 
obtained from the analysis of individual 
supergiant stars \citep{Kudritzki2024} and the 
stellar population synthesis studies of young 
stellar populations in SDSS \citep{Sextl2023} and 
TYPHOON \citep{Sextl2024} local star-forming galaxies. 
The Model\,B data point ($\rm [Z]_{\rm NGC300}\,=\,-0.2335$, 
the present-day predicted metallicity at a radius r\,=\,0.4R$_{25}$ as for all galaxies with a metallicity 
gradient in this plot) fits very nicely on the relationship.

\begin{figure}
  \centering
  \includegraphics[angle=0,scale=0.55]{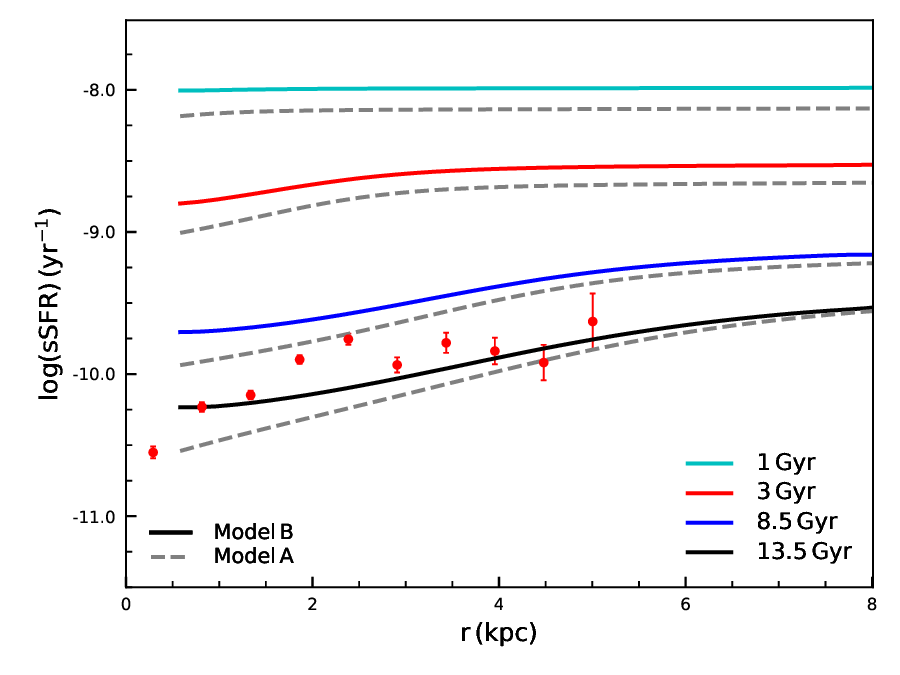}
    \caption{Cosmic time evolution of sSFR radial profiles. 
    Solid lines with different colours represent
    Model\,B at $1\,\rm Gyr$ 
    (cyan), $3\,\rm Gyr$ (red), $8.5\,\rm Gyr$ (blue), 
    and $13.5\,\rm Gyr$ (present-day, black). 
    The corresponding results by Model\,A  
    are shown as gray dashed lines. Red filled cycles 
    are the observed data, the same as those in the 
    right-top panel of Fig.\ref{fig:model_result}.
    }
  \label{fig:sSFR_evolution}
\end{figure}

A crucial aspect of galaxy evolution is the change of the
specific SFR. Figure\,\ref{fig:sSFR_evolution} plots 
the time evolution of sSFR radial profiles predicted by 
Model\,B (solid line) and Model\,A (dashed line). 
The values of sSFR are high at early times, 
and the radial profile is nearly flat. As NGC\,300 evolves, 
the initially flat sSFR gradient becomes steeper, due 
to the gas exhaustion of the inner region of the 
galaxy. (This trend is consistent with the 
model results obtained by \citet{Belfiore2019}).
In addition, compared to Model\,A, Model\,B 
predicted sSFR values are higher during the whole 
evolution history of NGC\,300. Radial gas inflows create
higher star formation activity.
Finally, the model\,B predicted present-day sSFR profile 
(i.e., at $13.5\,\rm Gyr$, black solid line) agrees 
well with the observed profile derived by using 
(${\rm FUV}-K$) colours \citep{MM2007}. We note that
the inclusion of radial gas inflows contributes 
to bringing the model predicted present-day sSFR 
profile much closer to the observations of NGC\,300 
(see the right-top panel of Fig.\,\ref{fig:model_result}), 
because the radial gas inflows provide more gas in 
the inner regions at later times.

\begin{figure}
  \centering
  \includegraphics[angle=0,scale=0.55]{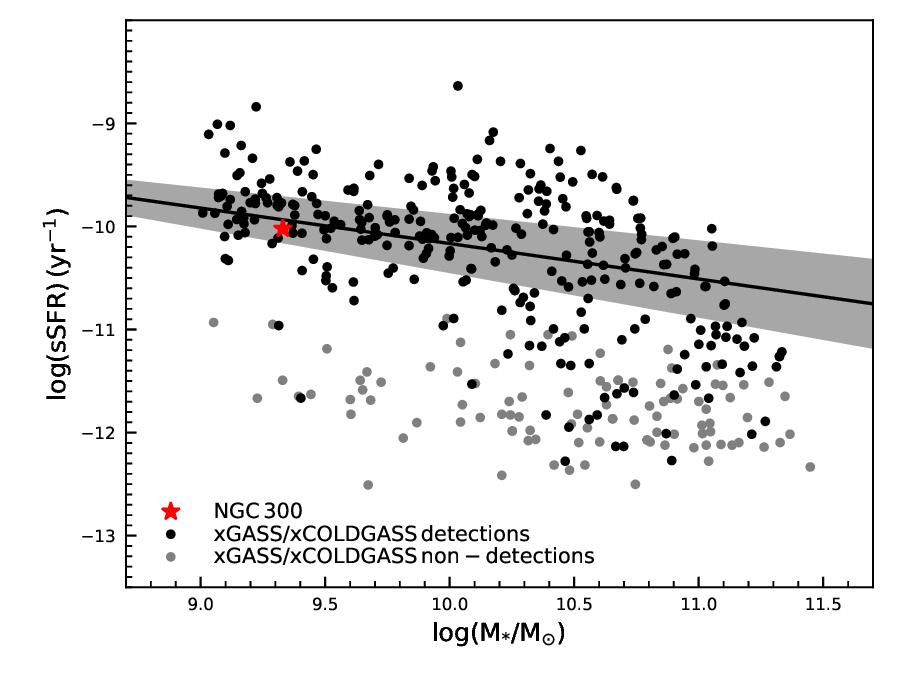}
    \caption{Specific SFR as a function of stellar mass, both 
    for the Model\,B predicted sSFR for NGC\,300 
    (red solid star) and the observed sSFR of xGASS 
    and xCOLDGASS \citep{Saintonge2017, Catinella2018}.
    The solid line refers to star-forming galaxy main sequence, 
    and the grey shaded region represents $1\sigma$ deviation 
    \citep[eq.2 in][]{Catinella2018}.
    Black filled circles refer to the observed SFR in nearby 
    galaxies from xGASS and xCOLDGASS detections, while 
    grey filled cycles denote non-detections in both 
    xGASS and xCOLDGASS.
}
\label{fig:M_sSFR}
\end{figure}

The sSFR of NGC\,300 predicted by Model\,B and Model\,A are $\rm log(sSFR/yr^{-1})\,=\,$-10.022 and -10.127, 
respectively. They agree well 
with the observed values (see Section\,\ref{sec:obv})
and fit also very nicely on the relationship with 
stellar mass obtained by xGASS and xCOLDGASS surveys
as shown in Figure.\,\ref{fig:M_sSFR}.
 
Figure\,\ref{fig:dif_r_Zgas} shows the time evolution of 
the radial metallicity profiles as predicted by Model\,B 
(solid line) and Model\,A (dashed line). 
The model starts at very low metallicity and with a strong 
metallicity gradient. Then, while the metallicity increases 
the gradient becomes flatter. This agrees well with one set 
of models and simulations \citep[e.g.,][]{BP2000, Molla2005, Pilkington2012, Minchev2018, Vincenzo2018, Acharyya2024}, 
but is in tension with several alternative approaches
\citep[e.g.,][]{Chiappini2001, Spitoni2013, Mott2013, Schonrich2017, Sharda2021, Graf2024}. 
In addition, at early times the radial gas inflows 
flatten the metallcity gradient, whereas at later times they
  steepen the gradient.
This is mainly due to the fact that, at early stages of 
evolution, the radial gas inflows dilute the metallicity. 
while at later times the radial gas inflows facilitate star 
formation and then enrich the metallicity.

As already demonstrated by Figure\,\ref{fig:model_result}, the
present-day metallicity of the young stellar population and ISM 
is reproduced well by our model. Now we also add the PNe studied 
by \citet{Stasinska2013} (see also \citealt{Magrini2016} for
construction of radial bins). With an average age of the 
PNe progenitors of 5\,Gyr, it is interesting to compare with 
the model metallicity at 8.5\,Gyr. Except for the two outer 
bins, the agreement is very good. As already discussed by
\citet{Stasinska2013}, the oxygen abundance of these outer 
PNe might be affected by nucleosynthesis in the AGB PNe 
progenitors.



\section{Conclusions}
\label{sec:sum}
NGC\,300 as a well studied isolated, bulge-less, and low-mass 
disc galaxy is ideally suited for an investigation of galaxy 
evolution. In this work, we build a bridge for NGC\,300 
between its observed properties and its evolution history 
by constructing a chemical evolution model. The main 
improvement of the model in this work is the inclusion of radial 
gas inflows. In addition, we extend the comparison and fit with the
observations by adding the results of extensive non-LTE spectroscopy
of very young massive stars.

Our model simultaneously reproduces the observed radial 
profiles of H{\sc i} gas mass surface density, SFR surface 
density, sSFR, as well as gas-phase, young stars and mean 
stellar metallicity and allows us to assess the effects 
of radial gas flows. While the radial flow velocity is very 
small, $\sim$ -0.1 km/s, it is sufficient to slightly steepen 
the H{\sc i} gas mass and SFR surface density profile. 
It flattens the radial profile of sSFR significantly
bringing it closer to the observations. The inclusion of radial
inflows leads to increasing star formation along the disc and
potentially helping to sustain star formation in local spiral arms. 

The effects of radial gas inflows on the present-day radial
metallicity distribution are small but the predicted influencing 
of the metallicity gradient when going back in time is 
significantly enhanced. The model predicted present-day 
metallicity fits nicely on the observed mass-metallicity 
relationship of star-forming galaxies. It
also agrees with the star-forming galaxy main sequence. 


\begin{acknowledgements}
We thank the anonymous referee for constructive 
comments and suggestions, which improved the quality of 
our work greatly.
Xiaoyu Kang and Fenghui Zhang are supported by the 
Basic Science Centre project of the National Natural 
Science Foundation (NSF) of China (No. 12288102),
the National Key R\&D Program of China with (Nos. 
2021YFA1600403 and 2021YFA1600400), the International 
Centre of Supernovae, Yunnan Key Laboratory (No. 202302AN360001),
the basic research program of Yunnan Province (No. 202401AT070142).
Rolf Kudritzki and Xiaoyu Kang acknowledge support by the Munich Excellence Cluster Origins and the Munich Instititute for Astro-, Particle and Biophysics (MIAPbP) both funded by the Deutsche Forschungsgemeinschaft (DFG, German Research Foundation) under the German Excellence Strategy EXC-2094 390783311. 
Xiaoyu Kang thanks Huanian Zhang in 
Huazhong University of Science and Technology 
for helpful suggestions during revision.

\end{acknowledgements}

%
%

\end{document}